# Application of Bayesian Methods for Age-dependent Reliability Analysis

**Robertas Alzbutas[1,2], Tomas Iešmantas[1]***

[1]Lithuanian Energy Institute, Breslaujos 3, LT-44403 Kaunas, Lithuania

[2]Kaunas University of Technology, Studentų 50, LT-51368 Kaunas, Lithuania

**Abstract**

In this paper authors present a general methodology for age-dependent reliability analysis of degrading or ageing systems, structures and components. The methodology is based on Bayesian methods and inference – its ability to incorporate prior information and on idea that ageing can be thought as age-dependent change of beliefs about reliability parameters (mainly failure rate), when change of belief occurs not just due to new failure data or other information which becomes available in time, but also it continuously changes due to flow of time and beliefs evolution.

The main objective of this paper is to present the clear way of how Bayesian methods can be applied by practitioners to deal with risk and reliability analysis considering ageing phenomena. The methodology describes step-by-step failure rate analysis of ageing systems: from the Bayesian model building to its verification and generalization with Bayesian model averaging which, as authors suggest in this paper, could serve as alternative for various goodness-of-fit assessment tools and as a universal tool to cope with various sources of uncertainty.

The proposed methodology is able to deal with sparse and rare failure events as is the case in electrical components, piping systems and various other systems with high reliability. In a case study of electrical instrumentation and control components the proposed methodology was applied to analyse age-dependent failure rate together with treatment of uncertainty due to age-dependent model selection.

## 1. Introduction

Reliability and safety of energy facilities, chemical factories, oil companies, etc. in many cases depends on their components reliability, which is mainly age-dependent. Component ageing is mainly caused by two impacts: operating conditions and technical inspection actions. Various ageing tests exist, see for example an excellent monograph by Lai and Xie[1], where they give lots of references about various aspects of aging identification. However, a case-specific consideration of ageing effects results into highly complex Probabilistic Risk Assessment (PRA) and the prevailing practice is to assume constant failure rate, sometimes even non conservative one. However, not taken into account at the time when safety margins are being estimated, ageing effect can cause failures or multiple damages at given non-standard operating conditions or breakdown situations.

The framework to deal with ageing in a coherent way depends on the type of data at hand. Statistical data can be represented as a pure failure sample, i.e.

---

it can be failure counts in consecutive (not necessarily equal) time periods, records of component state (failed or not) at specific times, or it might be evolution of component physical degradation characteristics, e.g. crack size.

There are a vast number of references, with models developed specifically to deal with the last type of data, known as degradation models. One of such comprehensive studies is a review of Singpurwalla *et al.*[2], where part of this paper is devoted to the stochastic diffusion–based state models and covariate induced hazard rate processes. Also, Yashi and Mantan[3] reviewed available literature on the likelihood construction for covariate induced hazard rate models when two cases are possible: unobserved and observed covariate processes.

As for the first type of statistical data, the closest paper to our research is written by Kelly and Smith[4], where they reviewed a state of the art of PRA with one of applications being related to ageing and valve leakage. However, they *a priori* assumed logit model and did not validated it except the comparison with the case where constant failure rate is assumed. In addition, some relevant papers are due to Colombo *et al.*[5], where authors present nonparametric estimation of time-dependent failure rate, or due to Ho[6], where semi-parametric family of bathtub shaped failure rates is analysed. Although these approaches offer a rich class of failure trend models, it also requires the larger samples of data. It is the price for the flexibility of these models.

Considering the way components function, the following division can be done: active (e.g. pumps, valves) and passive (e.g. heat exchangers, pipes, vessels, electrical cables, structures). For detailed explanation of the terms see IAEA Safety Glossary (Version 2.0, 2006).

The passive components are usually neglected or not modelled implicitly in the probabilistic reliability and risk assessment models of complex systems as having very low failure probability, but they could have an increasing contribution due to ageing effects. While safety importance of active systems and their ageing were recognised widely, in this paper the focus is more on complicated reliability data analysis of passive systems when the statistical information is provided as a sample of failure counts in consecutive periods of times.

Together with previously mentioned modelling complexity, which arises with ageing introduction into model, age-dependent reliability study requires more data for inferences to be valid. With regard to data, one basic issue is scattering of failure histories for passive components. Because of this scattering, reliability and risk model parameters, which are estimated from the raw data, have associated large uncertainty.

Usually, passive systems/components do not provide large samples of failure data. Classical statistics framework can be hardly applied to such problems due to small samples and asymptotic assumptions, like consistency or asymptotic normality of estimators.

On other hand, uncertainty, related to data scattering and the small samples might be reduced by prior information (if available) – experience of other similar facilities, subjective expert insights, etc. Then, by the use of available statistical data, prior knowledge can be revised by Bayes formula (Berger[7]).

Within Bayesian approach one can rely on multiple sources of evidence including: warranty data, customer research surveys, proving ground test data, etc. It also has the potential to systematically quantify and process "soft" evidence such as expert knowledge (Krivtso and Wasiloff[8]). However, elicitation of prior information is a very challenging task, and the usual way to proceed in Bayesian analysis is to employ so-called uninformative priors. The use of non-informative priors does not invalidate Bayesian analysis of small samples, though.

In addition to the ability to deal with sparse data, Bayesian methods are appropriate for use in PRA (Siu and Kelly[9]). Further, practical advantage of the Bayesian framework in PRA applications is that propagation of uncertainty through complex models is relatively simple. On the other hand, it is very difficult, and intractable in practical analysis, to propagate classical statistical confidence intervals through PRA models to estimate a confidence interval for a composite result of interest.

Despite of the advantages offered by Bayesian methods, practical applicability of it was very limited and generally confined by the use of so called conjugate prior distributions, which provides analytically tractable solutions just for quite unsophisticated applications.

However, the advent of Markov Chain Monte Carlo (MCMC) sampling has proliferated Bayesian inference throughout the world, across a wide array of disciplines (Kelly and Smith[10]). MCMC methods are a class of algorithms for sampling from probability distributions based on constructing a Markov chain that has the desired distribution as its equilibrium distribution (Gilks, Richardson and Spiegelhalter[11], Ching and Michael[12]). The state of the chain after a large number of steps is then used as a sample from the desired distribution. MCMC algorithms enabled development and application of highly complex Bayesian models. The freely available software package known as Bayesian inference Using Gibbs Sampling (WinBUGS) has been in the vanguard of this proliferation since the mid-1990s (Lunn, Andrew and Nicky[13]). Our calculations is performed by WinBUGS software to illustrate that outwardly complex MCMC algorithms could be quite easily used by reliability engineers to deal with aging and Bayesian models.

We construct our paper as an investigation of different Bayesian modelling steps in order to give for the reliability analyst and PRA practitioner a clearer view of how ageing phenomena can be easily incorporated within their tasks. The main contribution of this paper to the reliability assessment field is in purely Bayesian treatment of ageing phenomena, in time-dependant reliability modelling methodology and many-sided investigation of analysis steps.

The structure of paper is as follows: in section 2.1 the review of most commonly used ageing trend models is presented; further, in sections 2.2 to 2.4 we present general methodology of Bayesian model construction from prior distribution selection to the model checking with section 2.3 being more didactical on posterior construction for general trend functions. In practical part of the paper (starting form section 3.1) we treat ageing effect of instrumentation and control components. This case study (in section 3.2) is transformed into piecewise homogeneous Poisson regression model according to the theory presented in section 2.4. In the same section we further elaborate on practical issues of prior distribution selection by pointing some pitfalls of gamma prior distribution. Then, followed by short advises on the MCMC initialization and convergence in WinBUGS, we obtain posterior summaries of trend models under consideration. We make constructive critique of p-values based on chi-square discrepancy measure and propose alternative one in section 3.3. Then, before conclusions and final remarks we complete our paper with section 3.4 devoted to the illustration of Bayesian model averaging technique being handled by recent marginal likelihood calculation method which we easily implemented in WinBUGS.

## 2. Theoretical issues of age-dependent failure rate

### 2.1 Dynamic models for failure rate analysis

If the evidences show possible trend in statistical failure data, then one can consider trend model for failure rate. Several examples for failure rate trend function $\lambda(t)$ can be:

- piecewise constant $\lambda(t) = \lambda_i, t \in [\tau_i; \tau_{i+1}]$; (1)
- linear $\lambda(t) = \theta_1 + \theta_2 t$; (2)
- exponential (log-linear) $\ln \lambda(t) = \theta_1 + \theta_2 t$; (3)
- power-law (Weibull) $\lambda(t) = \theta_1 t^{\theta_2}$; (4)
- Xie and Lai model

$$\lambda(t) = \theta_1 \theta_2 (\theta_1 t)^{\theta_2 - 1} + \theta_3 \theta_4 (\theta_3 t)^{\theta_4 - 1}, 0 < \theta_2 < 1, \theta_4 > 1, \text{(Xie and Lai}[14]); \quad (5)$$

- generalized Makeham $\lambda(t) = \theta_1 e^{\theta_2 t} + \dfrac{\theta_3}{1 + \theta_4 t}$, (Lai and Xie[1]); (6)

where $\lambda(t)$ is age-dependent failure rate, $t$ - age covariate, $\theta$ - parameters which influence the shape of failure rate trend function.

Linear ageing is the mostly simple and obvious natural way to give a first-order approximation to changes in the failure rate, but it does seem to have a practical disadvantage. Wolford et al.[15] analyzed two data sets using several functional forms for $\lambda(t)$; one such analysis is reported by Atwood[16]. They found that a Bayesian posterior distribution for $\lambda(t)$ was approximately lognormal when a log-linear or power-law model was used for $\lambda(t)$, but this was not a case when a linear model was used. Apparently, the approximate log normality required a much larger data set when linear ageing was assumed in comparison to the case when power-law or exponential ageing were assumed.

Usually, the timing for failure rate consideration is divided into three distinct periods: burn-in period, useful life, wear-out period. For such general trend the linear, power-law or exponential distribution can't provide desirable fit to the data. Due to this reason, models that have ability to shape-up whole bathtub curve are needed and Xie & Lai or generalized Makeham trend models can be applied (Lai and Xie[1]). Notwithstanding flexibility of these models, it can be quite difficult to apply them by using frequentist framework, because due to number of parameters and lack of enough data to estimate them. Classical statistical methods are ill-suited for this situation, leading in such cases to excessively wide confidence intervals.

Some authors (Radionov[17], Okazaki and Aldemir[18], Radulovich, Veseley and Aldemin[19]) introduce a threshold of age at which ageing is assumed to begin. Then $\lambda(t)$ is assumed to be a constant before the threshold of age is attained, and to be increased according to one of the above formulas afterwards. The threshold is generally unknown, and must be estimated from the data. Thresholds cause difficulty in classical statistics, because the assumptions for the asymptotic theory of maximum likelihood estimation are typically violated. Therefore, it is difficult to quantify the uncertainty in the estimate of the threshold. However, even in this case the application of Bayesian modelling, for instance, using simulation package such as BUGS® (Lunn, Thomas and Best[13]), is still possible.

## 2.2　*Prior information in Bayesian modelling*

As noted in introduction section, Bayesian methods are capable to join various sources of information: statistical data, expert opinions, historical information, experience in quantification of uncertainty in similar systems or components, etc. These sources can be classified into subjective (e.g. expert opinions, etc.) and objective (statistical data).

To quantify subjective information, the need of subjective probability framework arises. The theory of subjective probability has been created to enable one to talk about probabilities when the frequency viewpoint for probability estimation does not apply. The main idea of subjective probability is to let the probability of an event to reflect the personal belief in the "chance" of the occurrence of the event (Berger[7]). In Bayesian theory, such subjective probability is expressed in terms of subjective prior distribution. More details about advantages and disadvantages of subjective probability can be found in (Chib, Clyde, Woodworth and Zaslavsky[20]).

Although subjective information, in the form of prior distribution, can be very useful, its elicitation is quite difficult and nontrivial task. Usual practice is to use expert opinion regarding some reliability characteristics (e.g. Coolen[21,22]); also meta – analysis can be used to quantify prior information in terms of distribution (Gelman *et al.*[23]). In addition, there exist in literature different prior distribution construction approach – data-based prior distribution construction, introduced by Guikema[24]. In this case, the basic idea is to divide the data sample into two parts – the first part is used to construct the prior distribution (by estimating the parameters with statistical methods), while the second part is used to obtain the likelihood function.

The basis of this approach is to use the historical data applying the method of moments, maximum likelihood or maximum entropy methods together with bootstrapping in order to obtain the estimates of prior distribution parameters. Then, the prior distribution together with the likelihood function, constructed for sample not used to obtain prior, yields a posterior distribution.

However, the situations when no useful prior information is available are quite frequent. In that case Bayesian modelling can be carried out by using non-informative prior or its proper approximation, which expresses prior ignorance about quantities of interest. This approach is often called objective Bayesian analysis and it is still ideally suited for small sample problems (Berger[25]).

Prevailing practice is to use Laplace uniform prior, which puts equal mass on whole real axis and is based on principle of insufficient reason or Jeffreys[26] invariant measure, based on Fisher information. In general, the use of non-informative prior distributions causes some problems in applications: such distributions are improper (i.e. they do not integrate to 1) and sometimes can yield improper posterior distribution; also Bayes factors, that are common quantities in models comparisons, cannot be calculated (Chib, Clyde, Woodworth and Zaslavsky[20]).

Proper prior distribution approximations can be used instead. Approximation of Laplace prior is a uniform distribution: in this case probability mass is distributed uniformly on some finite interval and gives no priority for any particular parameter value, as can be the case with Jeffreys prior approximation, which can bring bias into small sample analysis. In addition, uniform distribution can be though about as being a diffuse distribution, i.e. it closely approximates the state, when expert has no clue about which particular values of parameters could be given a priority. Diffuse priors can be gamma distribution with very small parameters (e.g. $\alpha = 0.001, \beta = 0.001$), normal distribution with very large variance, etc. But, as we will demonstrate in the application part, analyst must be very careful when he use diffuse priors (especially in the small sample case), because it can lead to incorrect inferences.

### 2.3 Application of Bayesian methods for age-dependent modelling

Ageing can be thought as age-dependent change of beliefs about reliability parameters (mainly failure rate). Change of beliefs occurs not just due to new failure data or other information (mentioned above) which becomes available in time, but also it continuously changes due to flow of time and beliefs evolution.

One of the difficulties of Bayesian inference is inability to deal with changes of age-dependent parameter as a continuous process. This problem partially can be overcome by considering ageing (or degradation) as step-wise process, which is constant in some period of time and has value jump in other period. Mathematically this can be expressed as a jump process:

$$d(t) = \sum_{i=1}^{N-1} 1_{\{t_i < t < t_{i+1}\}} d(t_i); \qquad (7)$$

where $d(t)$ is any model of characteristic (or reliability parameter) under consideration and constant $d(t_i)$ is value of characteristics at each time period $t_i$; $N$ – number of time intervals.

Model of characteristic $d(t)$ can have any functional form. It can be linear, Weibull, or some other form. Depending on adopted formula, $d(t)$ will be based on vector of parameters $\Theta = \{\theta_1, ..., \theta_m\}$:

$$d(t) = d(t, \Theta). \qquad (8)$$

If analyst considers more than one model, then indexation is used for different models, i.e. $d_i(t, \Theta_i)$, where $d_i$ denotes $i^{th}$ model with $\Theta_i$ vector of parameters.

Modelling conception introduced above allows interpreting distribution of parameters as age-dependent. If prior knowledge and beliefs about failure rate or other reliability parameter is represented by probability density distribution $\pi(\Theta)$ and statistical observations has likelihood $f(Y|d(t))$, where $Y=(y_1,...,y_N)$ is sample of observations, then, according to Bayes theorem, age-dependent beliefs about the reliability parameter is expressed as posterior distribution:

$$\pi(\Theta|Y,t) = \frac{\pi(\Theta)f(Y|d(t,\Theta))}{\int_\Omega \pi(\Theta)f(Y|d(t,\Theta))d\Theta}. \qquad (9)$$

Assume that parameters $\theta_1,...,\theta_m$ are a priori independent, then, according to definition of independent random variables, prior distribution of $\Theta$ can be expressed as:

$$\pi(\Theta) = \prod_{i=1}^{m}\pi_i(\theta_i), \qquad (10)$$

where $\pi_i(\theta_i), i=\overline{1,m}$ are priors for components of vector $\Theta$.

If data set contains $n$ statistical observations, then posterior distribution is represented as:

$$\pi(\Theta|Y,t) = \frac{\prod_{i=1}^{m}\pi_i(\theta_i)\prod_{j=1}^{n}f(y_j|d(t_j,\Theta))}{\int_\Omega \prod_{i=1}^{m}\pi_i(\theta_i)\prod_{j=1}^{n}f(y_j|d(t_j,\Theta))d\Theta}. \qquad (11)$$

In addition, usually it is the case when several trend models fits data almost equally well, i.e. possible set of "good" models contain more than one possibility

$$\mathbf{M} = (d_1(t,\Theta_1),...,d_r(t,\Theta_r)), \qquad (12)$$

where $d_i(t,\Theta_i), i=\overline{1,r}$ are models which were considered as having good fit.

In such circumstances uncertainty of modelling cannot be handled appropriately within classical statistical framework. As noticed by Hoeting[27], standard statistical practice ignores model uncertainty. Data analysts typically select a model from some class of models and then proceed as if the selected model had generated the data. This approach ignores the uncertainty in model selection, leading to over-confident inferences and decisions that are more risky than one thinks they are. As an alternative approach, models averaging is more correct because it takes into account a source of uncertainty that analyses based on model selection ignore (Kulinksaya, Morgenthaler and Staudte[28]). Also, according to Hoeting[27], Bayesian model averaging advantages include better average predictive performance than any single model that could be selected.

Then, let's denote $A(t)$ as failure rate averaged over set of models $M$. Considering our notation, posterior probability of averaged age-dependent failure rate can be represented as:

$$p(A(t)|Y) = \sum_{j=1}^{r} p(A(t)|Y, d_j(t,\Theta_j)) p(d_j(t,\Theta_j)|Y), \qquad (13)$$

where $p(d_j(t,\Theta_j)|Y)$ is posterior probability distribution of model $d_j(t,\Theta_j)$ given the set $M$ of available models; $p(A(t)|Y, d_j(t,\Theta_j))$ is posterior distribution of quantity $A(t)$ under model $d_j(t,\Theta_j)$. Posterior probability distribution for model $M_j$ is given by

$$p(M_j|Y) = \frac{p(Y|d_j(t,\Theta_j)) p(d_j(t,\Theta_j))}{\sum_{l=1}^{r} p(Y|d_l(t,\Theta_l)) p(d_l(t,\Theta_l))}, \qquad (14)$$

where $p(d_j(t,\Theta_j))$ is prior probability distribution of model $d_j(t,\Theta_j)$, $p(Y|d_j(t,\Theta_j))$ is marginal likelihood conditional on model $d_j(t,\Theta_j)$.

In the case of non-informative prior distribution, equal discrete probabilities can be assigned for each model $p(d_j(t,\Theta_j)) = \frac{1}{r}$ and posterior probability distribution for model $d_j(t,\Theta_j)$ becomes:

$$p(d_j(t,\Theta_j)|Y) = \frac{p(Y|d_j(t,\Theta_j))\frac{1}{r}}{\sum_{l=1}^{r} p(Y|d_l(t,\Theta_l))\frac{1}{r}} = \frac{p(Y|d_j(t,\Theta_j))}{\sum_{l=1}^{r} p(Y|d_l(t,\Theta_l))}, \forall j = \overline{1,r}. \qquad (15)$$

Even though, Bayesian Model Averaging (BMA) seems to have advantages over one-model-fitting, little work has been done in the engineering field to address this for model uncertainty. For example, Alvin et al.[29] used BMA to predict the vibration frequencies of a bracket component, Zhang and Mahadevan[30] applied it in fatigue reliability analysis on the butt welds of a steel bridge, and most recent work was done by Inseok Park et al.[31]. Those authors analyzed uncertainty of 4 finite element models for laser peening process. However, all of these works used relatively simple models with unsophisticated probabilistic assumptions and there was no need to adopt advanced probability sampling techniques such as Markov Chain Monte Carlo methods with validation of model selection.

### 2.4 *Model selection*

There are various techniques for model validation in Bayesian framework (Ntzoufras[32], Gelman, Meng and Stern[33], Dei and Rao[34]]. One of possible approaches to analyse model fitness is to use tail-area probability or as it is sometimes known, the posterior predictive p-value:

$$p = P(D(Y^{rep},\theta) > D(Y,\theta)|Y) = \iint I_{[D(Y^{rep},\theta) > D(Y,\theta)]} p(Y^{rep}|\theta) p(\theta|Y) dY^{rep} d\theta, \quad (16)$$

where $y^{rep}$ is the replicated data that could have been observed, or, to think predictively, as the data that would appear if the experiment that produced $y$ were replicated in future with the same model (Gelman, Meng and Stern[33]). Posterior p-value expresses the differences between statistical data and replicated. Rule of thumb is p-values close to 0.5 (Ntzoufras[32], Gelman and Meng[35]). $D(Y,\theta)$ is discrepancy measure and can have any functional form, e.g.:

$$D_1(Y;\theta) = \sqrt{E(Y-E(Y))^2} \text{ and } D_2(Y;\theta) = \sum \frac{(y_t - E[y_t|\theta])^2}{Var[y_t|\theta]}. \quad (17)$$

Chi-square statistics $D_2(Y;\theta)$ is quite popular among researchers; however as will be showed the use of just one discrepancy measure can be very misleading.
The use of discrepancy measures can be used to assess fitness of each model individually, i.e. rejection and acceptance of one model does not depend on other models.
Another possible way to analyse fitness of models is to use Deviance Information Criterion (DIC), which is already implemented in WinBUGS as inner function. DIC can be used to compare different models with each other. Spiegelhalter *et al.*[36] suggest the following rule of thumb: that models with DIC difference within the minimum value lower than two (2) deserve to be considered as equally well, while models with values ranging within 2-7 have considerably less support. DIC of i[th] model is defined as:

$$DIC_i = -2\ln(L(\Theta|Y,i)) + 2p_D; \quad (18)$$

where $p_D$ is the effective number of parameters (Spiegelhalter *et al.*[36]).
More information about Bayesian model selection can be found in (Ntzoufras[32], Dei and Rao[34]).

3. **Case study**

*3.1 Data representation*

Data set represents the failure and replacement dates of electrical instrumentation and control (I&C) components. The considered data is quite similar to the real operating experience data collected in French or German nuclear power plants (data were encoded and exact places where it was collected can't be identified). In particular, it is a large sample that represents one technological group of continuously operating components. The data set contains records from type "T" reactors, which are operated by a single utility with a single management philosophy. The components and composition of them in all reactors are similar (design, manufacturer, technology, etc.). In all type "T" reactors the components of type "A" are subjected for ageing effect during their operation in the environment with more stressful pressure and temperature. The scope of maintenance is the same for all components.

All data were collected during eleven years, from January 1, 1990 through December 31, 2000. The components in the sample do not all have same date of being put into service, and as a consequence do not have the same ages at the beginning and end of observation. This reason caused the expansion of age scale from 11 years to 15 years, i.e. at the beginning of the observation some units have been operating for several years already, and as a consequence their were older than 11 years. The failure counts were taken from a review of the maintenance data, so any reported date of failure is actually the date of the periodic test. A "critical" failure is one that causes the component to lose its safety function modelled for PRA.

There were 20 reactor units of type "T", each with 20 components of type "A". So, each year there would be 400 component-years except for the fact that some of the reactor units were commissioned before and after the start of the data collection (see Table 1). This caused differing cumulative operating times.

Failure rates, presented in Table 1, give the first impression about failure behaviour over time: failure rate increases in time showing system ageing effect. Also several statistical tests were performed for the ageing effect confirmation and were presented in the report of JRC Institute for Energy (Radionov[17]).

| Age, Years | Number of failures | Cumulative operating time, Years | Failure rate, 1/Year |
|---|---|---|---|
| 1 | 1 | 126.60 | 0.0079 |
| 2 | 1 | 171.62 | 0.0058 |
| 3 | 3 | 231.36 | 0.0130 |
| 4 | 1 | 314.80 | 0.0032 |
| 5 | 10 | 396.60 | 0.0252 |
| 6 | 8 | 400.00 | 0.0200 |
| 7 | 16 | 396.76 | 0.0403 |
| 8 | 11 | 380.00 | 0.0289 |
| 9 | 12 | 363.34 | 0.0330 |
| 10 | 8 | 336.73 | 0.0238 |
| 11 | 16 | 281.68 | 0.0568 |
| 12 | 9 | 273.42 | 0.0329 |
| 13 | 10 | 288.44 | 0.0347 |
| 14 | 16 | 168.58 | 0.0949 |
| 15 | 15 | 85.16 | 0.1761 |

Table 1. Failure data of I&C components under consideration

### 3.2 Bayesian model for piecewise homogeneous Poisson count data

In this analysis, failure rates are considered as constant values in each year, but at every year this value jumps at the value which can be calculated from linear, Weibull or other model.

Consider as the model for the failure rate $\{\lambda(t); t \geq 0\}$ a jump process structure described above:

$$\lambda(t) = \sum_{i=1}^{N} 1_{\{t_i < t \leq t_{i+1}\}} \lambda_i. \tag{19}$$

In each year period failures occurs as homogeneous Poisson process but with different failure rate parameter $\lambda_i, i = 1, 2, ..., 15$. In every time period (which in this case is equal to one year) equipment was in operation for $\tau_i$ time (operating time). Denote number of failure that occurred in one year as $N_i$. Probability of $N_i$ failure can be expressed as:

$$P(N_i = k) = \frac{e^{-\lambda_i \tau_i} (\tau_i \lambda_i)^k}{k!}. \tag{20}$$

Likelihood function, that contains all information obtained from data, is:

$$L(\Theta) = P(Y | \Theta) = \prod_{i=1}^{n} \exp\{-\lambda(t_i, \Theta) \tau_i\} \frac{(\lambda(t_i, \Theta) \tau_i)^{N_i}}{N_i!}. \tag{21}$$

### 3.2.1. Selection of prior distribution

Since in data source (Radionov[17]) there is no available information about which particular I&C components were under observation, prior distribution for parameters of failure trend function is chosen as diffuse distribution.

As already mentioned in Section 2.2, to express diffuse knowledge about model parameters, one can choose to use uniform, gamma, normal distributions, etc. It is our experience that gamma distribution with small parameters (which is the usual way to obtain diffuse prior) can lead to incorrect estimates. This occurs due to nature of gamma distribution – all its mass is concentrated close to zero (Figure 1.).

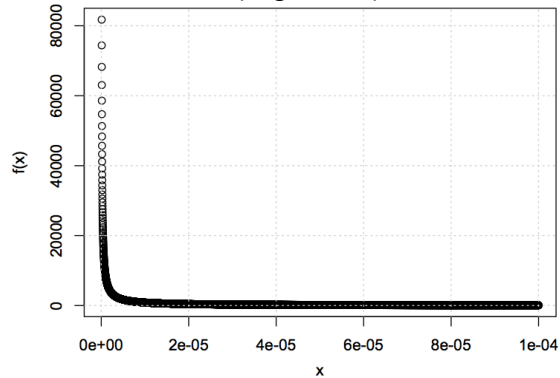

Figure 1. Gamma distribution with parameters $\alpha = \beta = 0.001$.

Due to this high concentration, prior gamma distribution sometimes is able to pull parameter estimates towards zero. This effect misrepresents the real underlying failure trend and causes to make overly optimistic decisions.

Assume linear model $\lambda(t) = a + bt$ and in one case all priors are gamma with parameters $\alpha = \beta = 0.001$, while in second case all priors are uniform distributions on interval $[0,100]$. Resulting posterior distribution for parameter $a$ are highly biased towards zero under gamma prior distribution (Figure 2).

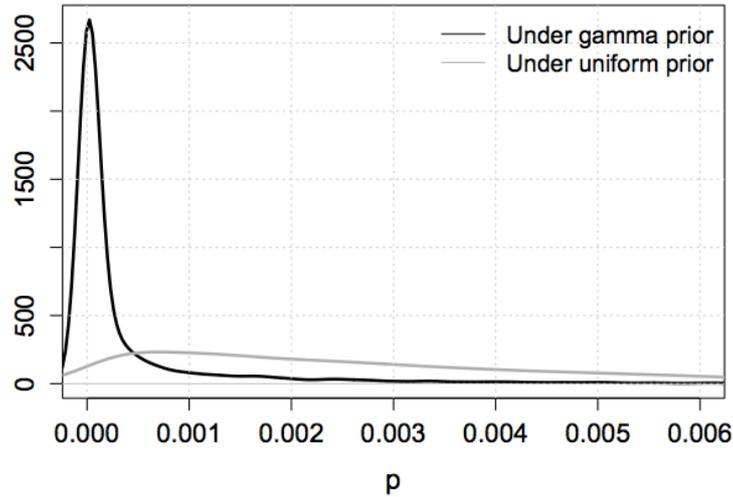

Figure 2. Influence of prior distributions for posterior inference.

Due to this observation, we confined ourselves with uniform prior distributions for all models and all parameters. Having this, prior distribution can be generally expressed as follows:

$$\pi(\Theta) = \prod_{i=1}^{m} \frac{1}{range(D_i)},$$

where range is a length of the interval $D_i$ with $i^{th}$ parameter is defined.

If the parameter is defined on positive part of real axis, it is not necessary to define prior in the same range i.e. on infinite interval. It is usually sufficient to choose "big enough" real value instead of infinity. Of course, posterior distribution has to be inspected and if at least one distribution is found to be truncated, one needs to extend the interval of prior distribution. Numerical experiments and constraints of parameters led to the following intervals of uniform prior distributions:

| Failure rate model | Parameters and their ranges for prior distributions |
|---|---|
| Linear | $(\theta_1,\theta_2) \in [0,100] \times [0,100]$ |
| Log-linear | $(\theta_1,\theta_2) \in [-100,100] \times [-100,100]$ |
| Power-law | $(\theta_1,\theta_2) \in [0,100] \times [-100,100]$ |
| Xie & Lai | $(\theta_1,\theta_2,\theta_3,\theta_4) \in [0,100] \times [0,1] \times [0,100] \times [1,100]$ |
| Generalized Makeham | $(\theta_1,\theta_2,\theta_3,\theta_4) \in [0,100] \times [-100,100] \times [0,100] \times [0,100]$ |

Table 2. Prior distributions of parameters definition ranges

### 3.3.2 MCMC convergence assessment in WinBUGS

Although, there exists quite comprehensive treatment of convergence monitoring in WinBUGS (see for example a book of Ntzoufras[32] or Kelly and Smith[37]), we think that several issues has to be addressed here in order to help practitioner to use WinBUGS more efficiently.

The first thing that we want to stress is initialization of Markov chains in BUGS software. There exist several options: one can choose initial values of chains by inner BUGS algorithm, or another one can provide them as a data list. In our case, enabling to initialize Markov chain automatically usually led to an error, that BUGS cannot initialize algorithm. This error was observed for gamma and for uniform prior distributions, although it was more often in former case. Interestingly, setting initial values by hand, quite robust reaction was observed regarding the accuracy of initialization, i.e. initial values does not have to be very close to the posterior expectation values in order to run algorithm without crashing of software.

Second point that we want to state is that gamma prior distribution, when algorithm successfully initialized, resulted to quite long calculations of chain values. As for uniform prior, software performed quite smoothly and the convergence was achieved almost immediately.

### 3.3.3 Posterior summaries

Having guaranteed convergence of the Markov chain, one has to decide (since no rigorous calculations can be done) on how long to run chains in order to obtain posterior summaries. These summaries in WinBUGS can be obtained very easily without any additional burden. On the other hand, such automation of calculation of summaries sometimes can get a bit restrictive, since just expectation, standard deviation and quantiles are reported. But if one wishes to employ other than squared error loss function (which corresponds to the posterior mean), then WinBUGS *Coda*[13] should be called for and values of Markov chain for specific parameter should be transferred and analysed outside the BUGS. For example this could happen if asymmetric loss function is used. Since our work is not directed to the analysis of various estimators, we confined ourselves to the posterior expectation as appropriate posterior summaries.

As mentioned at the beginning of this paper, 5 trend models (Figure 3) of failure rate were considered. Linear, exponential and power models represent class of trends which is common in ageing analysis and Makeham and Xie & Lai models represents more flexible bathtub trend class. We excluded constant failure rate model because ageing effect of considered data has been already validated in other analysis (Radionov[17]).

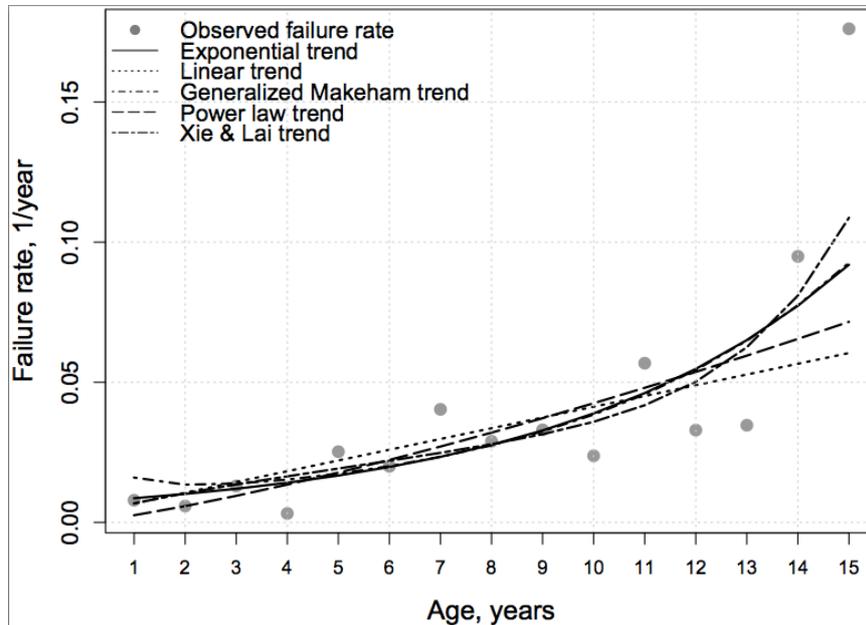

Figure 3. Comparative representations of fitted trend models

The posterior summaries are being reported in the following table.

| Failure rate model | Posterior expectations of parameters | | | |
|---|---|---|---|---|
| Linear: $\exp(\theta_1 + \theta_2 t)$ | 0.0073 | 0.1704 | | |
| Log-linear: $\theta_1 + \theta_2 t$ | 0.0030 | 0.0038 | | |
| Power-law: $\theta_1 t^{\theta_2}$ | 0.0025 | 1.2710 | | |
| Xie & Lai: $\theta_1 \theta_2 (\theta_1 t)^{\theta_2 - 1} + \theta_3 \theta_4 (\theta_3 t)^{\theta_4 - 1}$ | 0.0700 | 5.7060 | 0.0113 | 0.8932 |
| Generalized Makeham: $\theta_1 e^{\theta_2 t} + \dfrac{\theta_3}{1 + \theta_4 t}$ | 0.7715 | 1510.0 | 0.0046 | 0.2118 |

Table 3. Posterior summaries for considered trend models

### 3.3  Model screening

Estimated posterior p-values for different failure rate models are in Table 4:

|  | Linear | Exponential | Power law | Gen. Makeham | Xie & Lai |
|---|---|---|---|---|---|
| $p(D_1)$ | 0.5458 | 0.6333 | 0.7356 | 0.6178 | 0.7006 |
| $p(D_2)$ | 0.0042 | 0.0278 | 0.0084 | 0.0306 | 0.0110 |

Table 4. Posterior p-values for different failure rate models

As can be seen from posterior p-values $p(D_2)$ presented in Table 4, in this case (using chi-square measure) none of proposed trend models of failure rate gives good enough fit and all models should be rejected. However, p-values $p(D_1)$ shows satisfactory discrimination abilities – linear, generalized Makeham and exponential trend models can be interpreted as better fit than Xie & Lai and power low failure rate trend models. Validity of p-values $p(D_1)$ is supported by a visual inspection of replicated and observed number of failures (Figure 4).

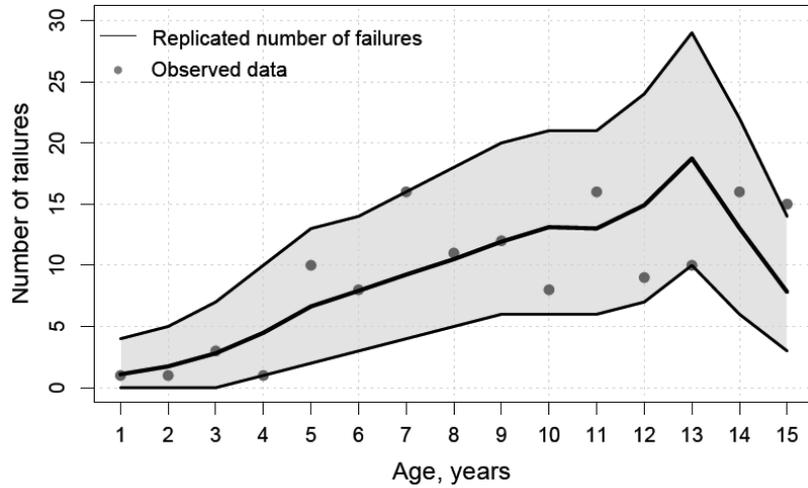

Figure 4. Replicated number of failures from exponential model compared to observed data

It is worth to take notice of chi-square discrepancy measures $p(D_2)$ inability to assess model goodness-of-fit, even though graphical investigation shows quite tolerable fitness. Even though standard deviation measure seems to work, but it might be that applied to another data sample it fails, as is the case with chi-square measure in this case. This leads to the conclusion that discrepancy measures (and as a consequence, posterior predictive p values) does not provide automatic model assessment tool for practitioners.

It is well known that more complex curves will fit data more precisely, but fitness of very complex models can lead to over fitting (e.g. perfect fitness can be achieved by splines, but this apparently leads to nonsensical inference).

Nevertheless, this obscurity can be solved by using DIC measure. This criterion naturally adopts Occam's razor principle, because it incorporates penalty - the effective number of parameters: more complex models will be penalized more severely. DIC values for all models under consideration are presented in Table 5.

| Model | Linear | Exponential | Power law | Gen. Makeham | Xie & Lai |
|-------|--------|-------------|-----------|--------------|-----------|
| DIC   | 91.39  | 86.48       | 88.42     | 94           | 88        |

Table 5. Values of Deviance Information Criterion

As can be seen from DIC values, exponential model shows best fit. Also, Xie & Lai and power law model can be accepted.

Two measures of fitness – discrepancy measure and DIC – shows different results and unambiguous answer cannot be given. Preference to one model over another can lead to too pessimistic or optimistic predictions of ageing phenomena behaviour. Such uncertainty related to the selection of model for further use has to be quantified to make sure that applications of model will not be influenced on incorrect choice of trend. Such quantification will be demonstrated in further analysis where Bayesian model averaging (BMA) will be applied.

### *3.4  Bayesian averaging for age-dependent failures*

As was concluded previously, discrepancy measure and DIC gave quite ambiguous results; subsets of models, selected by these criteria are not exactly the same. In practice, usual decision is to adopt just one model, but as we have already mentioned in theoretical part, this could lead to overoptimistic results if model uncertainty is not incorporated into modelling process.

In this part of paper application of Bayesian model averaging to analyze age-dependent failures will be demonstrated. We will perform averaging procedure for all models that were considered in this paper. To be able to average over the set M of models, probabilities of each model has to be obtained by calculating marginal likelihoods with WinBUGS software. Calculation of marginal likelihoods is not a trivial task, especially with WinBUGS, where user cannot control which MCMC method to use. However, quite recently Friel and Pettitt[38] proposed a method, when marginal likelihood can be estimated via power posteriors, defined as follows:

$$\pi_s(\Theta|Y,t) \propto f^s(Y|d(t,\Theta))\pi(\Theta), \ s \in [0,1].$$

It can be proved, that:

$$\log p(Y|d(t,\Theta)) = \int_0^1 E_{\Theta|Y,s}\left[\log f(Y|d(t,\Theta))\right]ds,$$

where expectation is taken over the power posterior. Then this integral can be approximated by a trapezoidal rule like this:

$$\log p(Y|d(t,\Theta)) \approx \frac{1}{2}\sum_{i=0}^{n-1}(s_{i+1}-s_i)\left(E_{\Theta|Y,s_{i+1}}\left[\log f(Y|d(t,\Theta))\right] + E_{\Theta|Y,s_i}\left[\log f(Y|d(t,\Theta))\right]\right).$$

To be able to obtain power posteriors in WinBUGS, one needs to define new sampling distribution with additional power parameter $s$ (for more details see WinBUGS manual and the appendix, wher the program code for exponential trend model case is presented). Then calculate expectations of log-likelihood for original model with regard to power posterior:

$$E_{\Theta|Y,s_i}\left[\log f(Y|d(t,\Theta))\right].$$

Suppose for a moment, that we have failure rates for 14 years and want to predict for the next one. We will apply BMA for this exercise. Obtained probabilities for each model are presented in Table 6.

| | Linear | Exponential | Power law | Gen. Makeham | Xie & Lai |
|---|---|---|---|---|---|
| $p(d_j(t,\Theta_j)|Y)$ | 0.107 | 0.422 | 0.108 | 0.216 | 0.147 |

Table 6. List of probabilities of analysed models

Calculated probabilities partially justifies assessments made by DIC and do not confirm conclusions based on discrepancy measures. Predictions together with 95% confidence regions are presented in Figure 5.

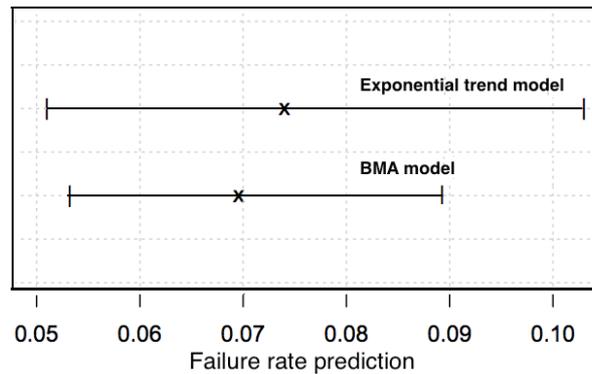

Figure 5. Bayesian predictive confidence intervals under exponential and BMA models

Differences in confidence intervals suggest that the statistical information about future values is not the same for exponential and BMA models. The smaller confidence interval more information is carried by the model from which it is obtained. The manifestation of shrinkage effect in BMA case is due to aggregation of information over the set of different models. We see, that BMA is superior to single model (exponential in our case) in reliability prediction task.

Having this, authors think that Bayesian averaging procedure can be good alternative to various goodness-of-fit approaches since it prevents decision maker of exclusion of models, which have good fit and could lead to reasonable posterior inferences. Also the information aggregation inherited by BMA approach can be advantageous over the single parametric model in component reliability prediction.

## 4. Conclusions and final remarks

In this paper authors presented general methodology of Bayesian methods application for age-dependent analysis. It was showed that this methodology is able to deal with disperse and small data amount along with multiple parameter set (Makeham and Xie & Lai trend models). As an illustrative example, the proposed methodology was applied for ageing analysis of electrical I&C components. This application was carried in terms of piecewise homogeneous Poisson model with several failure trends.

When fitting and screening various trend models, it was noticed that none of model selection approaches can give unambiguous answer. P-values can be quite misleading and can either show no discriminatory abilities (as in case of chi-square p-value) or can suggest more than one model as having good fit (as in case of standard deviation p-value).

Deviance information criteria can also suggest more than one model (and not necessarily the same one as p-value criteria). That's why there is a high chance to omit model which can also lead to satisfactory results. Thus, model selection should be performed very carefully. It is worth to mention, that other model selection and validation criteria (such as Bayesian information criteria, Bayesian factors, etc. not described or used in this paper) can also suffer from such shortcomings.

To evade the drawbacks of model selection and validation tools, Bayesian posterior model averaging procedure were performed for whole set of models, which were analysed in this paper. Such averaging over set of selected trends finally results to better predictive performance, because, due to shrinkage effect inherited by BMA approach, averaged future failure rates will not be underestimated in terms of their uncertainties. Notwithstanding all the advantages of Bayesian model averaging, this approach also undergoes some problems: BMA cannot deal with infinite set of models and when one chooses finite set of them the best one can be not included in this set; it also fails to "emit the alert signal" when all models fits data very poorly and so averaging will not result to better performance.

This paper and its results can be used as groundwork for further assessment of ageing components. Its generality and idea, that ageing or degradation can be thought as age-dependent change of beliefs about reliability parameters, allows analysis of wide spectrum of problems - it can be stochastic behaviour of crack growth (in this case characteristic $d(t)$ of interest would be crack growth rate), it can be degradation modelling as transitions through Markovian states ($d(t)$ could be transition rates, time-homogeneous or time-inhomogeneous, between degradation states), etc.

## Acknowledgement

This research was partially funded by the grant (No. ATE-10/2010) from the Lithuanian Research Council.

Appendix A

We present here a WinBUGS program script for calculation of power marginals for exponential model. Obtained posterior expectations of the variable *logpower* are then used together with trapezoidal rule to estimate the marginal likelihood value.

```
model{
 #Poisson model with exponential trend line, p - parameters
  for(i in 1:N){
     x[i]~dpois(rate[i])
     rate[i]<-lambda[i]*time[i]
     lambda[i]<-exp(p[1]+p[2]*T[i])
  }
 # Construction of power likelihoods
 #For each ith power, we construct a corresponding likelihood model by "zero trick"
  for(i in 1:100){
    #For each data point obtain values of power-likelihood
    for(j in 1:N){
       zeros[i,j]<-0
       zeros[i,j]~dpois(phi[i,j])
       # pow(0.01*I,3) is chosen to obtain uneven distribution of powers
       phi[i,j]<-  pow(0.01*i,3)*(rate_p[i,j]-x[j]*log(rate_p[i,j])+logfact(x[j]))
       rate_p[i,j]<-lambda1[i,j]*time[j]
       lambda1[i,j]<-exp(p1[i,1]+p1[i,2]*T[j])
    }
  }
 #Calculate power posterior over entire data sample
  for(i in 1:100){
     logpower[i]<- -sum(phi[i,1:N])/pow(0.01*i,3)
  }

 #Exponential model priors
  p[1]~dunif(-10,10)
  p[2]~dunif(-10,10)
 #Priors for power-likelihood
  for(i in 1:100){
     p1[i,1]~dunif(-1000,100)
     p1[i,2]~dunif(-1000,100)
  }
}
```